\documentclass[prl,longbibliography,twocolumn,nopacs,preprintnumbers,notitlepage,amsmath,amssymb,superscriptaddress]{revtex4-2}

\usepackage{epsfig}

\usepackage[colorlinks=true,urlcolor=blue,citecolor=blue,linkcolor=blue]{hyperref}
\hypersetup{breaklinks=true}

\usepackage{graphicx}  
\usepackage{dcolumn}   
\usepackage{bm}        
\usepackage{amsmath}
\usepackage{amsfonts}
\usepackage{bbm}
\usepackage{float}
\usepackage[caption=false]{subfig}
\usepackage{enumerate}
\usepackage{tikz}
\usepgflibrary{arrows.meta}
\usepackage{physics}
\usepackage{mathrsfs}
\usepackage[normalem]{ulem}
\usepackage{graphicx}
\usepackage[caption=false]{subfig}
\usepackage{makecell}

\usepackage{lipsum}
\begin{document}

\title{Nematic Bogoliubov Fermi surfaces from magnetic toroidal order in FeSe$_{1-x}$S$_x$}
\author{Hao Wu}
    \thanks{These authors contributed equally.}
    \author{Adil Amin}%
    \thanks{These authors contributed equally.}
\author{Yue Yu} 
\author{Daniel F. Agterberg}
\affiliation{Department of Physics, University of Wisconsin--Milwaukee, Milwaukee, Wisconsin 53201, USA}
\date{\today}

\begin{abstract} 
Recently it has been shown experimentally that the superconducting state of FeSe$_{1-x}$S$_x$ exhibits Bogoliubov Fermi surfaces for $x>0.17$. These Bogoliubov Fermi surfaces appear together with broken time-reversal symmetry and surprisingly demonstrate nematic behavior in a structurally tetragonal phase. Here, through a comprehensive analysis that deduces the structure of Bogoliubov Fermi surfaces from all symmetry-allowed types of translation invariant broken time-reversal symmetry,  we find that the origin of nematic Bogoliubov Fermi surfaces is due to magnetic toroidal order that belongs to an $E_u$ irreducible representation of the $D_{4h}$ point group. We further show that this magnetic toroidal order appears as a consequence of either static  N\'{e}el antiferromagnetic order or due to the formation of a spontaneous pair density wave superconducting order. Finally, we reveal that independent of the presence of Bogoliubov Fermi surfaces, supercurrents will induce N\'{e}el magnetic order in many Fe-based superconductors.
\end{abstract}

\maketitle

{\it Introduction.}---It has recently been realized that in superconductors when time-reversal symmetry is broken, the gap is either fully gapped or has Bogoliubov Fermi surfaces (BFSs)~\cite{agterberg2017bogoliubov, brydon2018bogoliubov}. BFSs are a type of nodal state in which usually expected line nodes or point nodes become surface nodes. In addition to altering the expected thermodynamic and transport response of superconductors, BFSs also reveal a generic weak-coupling instability into a broken-inversion state~\cite{oh2020instability, tamura2020electronic} and the emergence of the spatially uniform odd-frequency  pairing~\cite{miki2021odd}. The experimental discovery of nematic BFSs in Fe-based superconductors represents a welcome platform to better understand this nodal state.

Among the iron-based superconductors~\cite{kamihara2006iron}, iron selenide (FeSe), with the simplest crystal structure and chemical composition, has attracted much attention~\cite{hsu2008superconductivity,shibauchi2020exotic,kreisel2020remarkable,coldea2021electronic}. It adopts a tetragonal structure at room temperature. Upon cooling, there is a structural transition from tetragonal to orthorhombic (nematic) phase at low temperature under ambient pressure~\cite{margadonna2008crystal,millican2009pressure,mcqueen2009tetragonal}. 
However, the nematicity in the isovalently substituted FeSe$_{1-x}$S$_x$~\cite{mizuguchi2009substitution,guo2014superconductivity} is strongly suppressed with sulfur (S) doping~\cite{watson2015suppression,moore2015evolution}, and it is completely suppressed at a sulfur content $x \approx 0.17$, indicating a nematic quantum critical point~\cite{hosoi2016nematic}.

In the tetragonal phase of FeSe$_{1-x}$S$_x$, there are experimental signatures for Fermi surfaces in the superconducting state. In particular, these BFSs reveal themselves through a large residual density of states (DOS) at the chemical potential that has been observed in the superconducting state through specific heat, thermal conductivity, and scanning tunneling spectroscopy (STS) measurements~\cite{sato2018abrupt,hanaguri2018two, mizukami2023unusual}.  Furthermore, evidence for broken time-reversal symmetry in FeSe$_{1-x}$S$_x$ by muon spin relaxation ($\mu$SR) measurements has also appeared~\cite{matsuura2023two}.
Time-reversal symmetry breaking (TRSB) is known to play a central role in stabilizing BFSs~\cite{agterberg2017bogoliubov, bzduvsek2017robust, link2020bogoliubov, Sumita:2019, oh2020instability, tamura2020electronic, miki2021odd, menke2019bogoliubov, timm2017inflated, setty2020topological, setty2020bogoliubov}.  TRSB, together with the preservation of parity symmetry, has been argued to give rise to topologically protected BFSs in FeSe$_{1-x}$S$_x$~\cite{matsuura2023two,setty2020topological,setty2020bogoliubov}, however, the detailed microscopic mechanism for these BFSs remains unclear.  

More recently, laser-based angle-resolved photoemission spectroscopy (laser ARPES) has directly observed a BFS in the tetragonal phase of FeSe$_{1-x}$S$_x$~\cite{nagashima2022discovery}. Surprisingly, this BFS has nematic symmetry, even though FeSe$_{1-x}$S$_x$ is structurally tetragonal. Here, we use a symmetry-based analysis of BFSs to show that the origin of such a nematic BFS is a magnetic toroidal (MT) order~\cite{dubovik1990toroid, fiebig2005revival, ederer2007towards,spaldin2008toroidal,kopaev2009toroidal, hayami2014toroidal,toledano2015primary, gnewuch2019fourth, hayami2018microscopic, hayami2018classification, hayami2022nonlinear, yatsushiro2022analysis} with an $E_u$ irreducible representation (irrep) of the $D_{4h}$ point group. This order differs from previous suggestions \cite{matsuura2023two, setty2020topological,setty2020bogoliubov} of a parity-preserving TRSB order in that the magnetic toroidal order breaks both parity symmetry ($\mathcal{P}$) and time-reversal symmetry ($\mathcal{T}$) while preserving the product of the two. With this $E_u$ MT order, we find that both the shape of the BFS and the momentum dependence of the minimum of the quasiparticle excitation energy agree with the laser ARPES results~\cite{nagashima2022discovery}. 

We further suggest two possible ways to realize $E_u$ MT order: the appearance of static N\'{e}el (checkerboard) antiferromagnetic (AFM) order with moments aligned in-plane, or the formation of a spontaneous pair density wave (PDW) superconducting order with an order parameter $\psi=\psi_0 e^{i 
\boldsymbol{q} \cdot \boldsymbol{r}}$~\cite{liu2023pair}. Experimental evidence exists for checkerboard AFM fluctuations in three-dimensional (3D) FeSe~\cite{wang2016magnetic, baum2019frustrated}. Checkerboard AFM order has been reported in two-dimensional (2D) FeSe~\cite{ qiao2020fingerprint}. In addition, a recent proposal showed that the checkerboard AFM order together with the substrate provides a realistic explanation for broken time-reversal symmetry seen in many related compounds, suggesting that checkerboard AFM order is likely to appear in these states~\cite{mazin2023induced}. Like magnetic toroidal order, both the AFM order and PDW order are $\mathcal{P}$ breaking, $\mathcal{T}$ breaking, and $\mathcal{PT}$ preserving. Furthermore, the momentum $\boldsymbol{q}$ of the PDW order \cite{Agterberg:2015} and the AFM order both belong to the same $E_u$ irrep of the $D_{4h}$ point group. This implies that in the superconducting state, these two orders are coupled. More generally, this coupling implies that an in-plane supercurrent in any P4/nmm Fe-based superconductors will generically induce AFM order with in-plane moments.

{\it Symmetry analysis of BFSs.}---All existing mechanisms for the appearance of BFSs require a translation invariant TRSB order \cite{agterberg2017bogoliubov, bzduvsek2017robust, link2020bogoliubov, Sumita:2019,setty2020bogoliubov}.  Here we carry out a symmetry analysis of the role of this TRSB on the Bogoliubov quasiparticle spectrum.  We explicitly consider the single-band limit in this analysis. While Fe-based superconductors are multiband systems~\cite{coldea2021electronic}, the observed low $T_c$ in FeSe$_{1-x}$S$_x$  suggests that interband pairing interactions will not significantly alter the Bogoliubov quasiparticle spectrum, so a single-band analysis should suffice.  

In our analysis, the key interaction is due to the TRSB, which alters the normal state Hamiltonian. This interaction can be external or induced by the broken time-reversal symmetry in the superconducting state \cite{agterberg2017bogoliubov, Kanasugi:2022}, the origin of this term is not essential for our analysis of the quasiparticle spectrum. For a single-band, with two pseudospin degrees of freedom, the normal state Hamiltonian with TRSB takes the general form 
\begin{equation}
H_N=[\xi_+({\bf k})+\xi_-({\bf k})]\sigma_0+{\bf h}({\bf k})\cdot\boldsymbol{\sigma},
\label{singleband}
\end{equation}
where the Pauli matrices $\sigma_i$ describe the spin degrees of freedom, $\xi_+({-\bf k})=\xi_+({\bf k}$),  $\xi_-({-\bf k})=-\xi_-({\bf k})$, and ${\bf h}(-{\bf k})={\bf h}({\bf k})$.  The interaction $\xi_-({\bf k})$ describes parity-odd time-reversal symmetry breaking while  ${\bf h}({\bf k})$ describes parity-even time-reversal symmetry breaking. These interactions can further be classified by which irrep of the $D_{4h}$ point group they belong to. Furthermore, since the observed BFSs are near the $\Gamma$ point, we carry out a power series expansion in momentum for $\xi_-({\bf k})$ and  ${\bf h}({\bf k})$ consistent with the symmetry properties defined by the corresponding irrep. These interactions, together with the resulting BFSs are given in Table~\ref{structures}. For simplicity, we take $\xi_+({\bf k})= \frac{\hbar^2 (k_x^2+k_y^2)}{2 m}$. For FeSe$_{1-x}$S$_x$, the normal state dispersion near the $\Gamma$-point has a weak $k_z$ dependence and also has a fourfold in-plane anisotropy, but this will not qualitatively change the results. 

To describe the superconducting state, we assume spin-singlet pairing with a fourfold anisotropic $s$-wave gap given by $\psi({\bf k})=\Delta_0+\Delta_4 \cos (4\theta)$ with $\Delta_0>0$, $\Delta_4<0$, and $\Delta_0>\left|\Delta_4\right|$, where $\theta$ is the polar angle for the in-plane momentum $\bf k$. Such a gap function is qualitatively consistent with the thermal conductivity measurements~\cite{sato2018abrupt} and the spectroscopic-imaging scanning tunneling microscopy measurements~\cite{hanaguri2018two} on  FeSe$_{1-x}$S$_x$.

\begin{figure} 
\centering
\includegraphics[height=5.4cm]{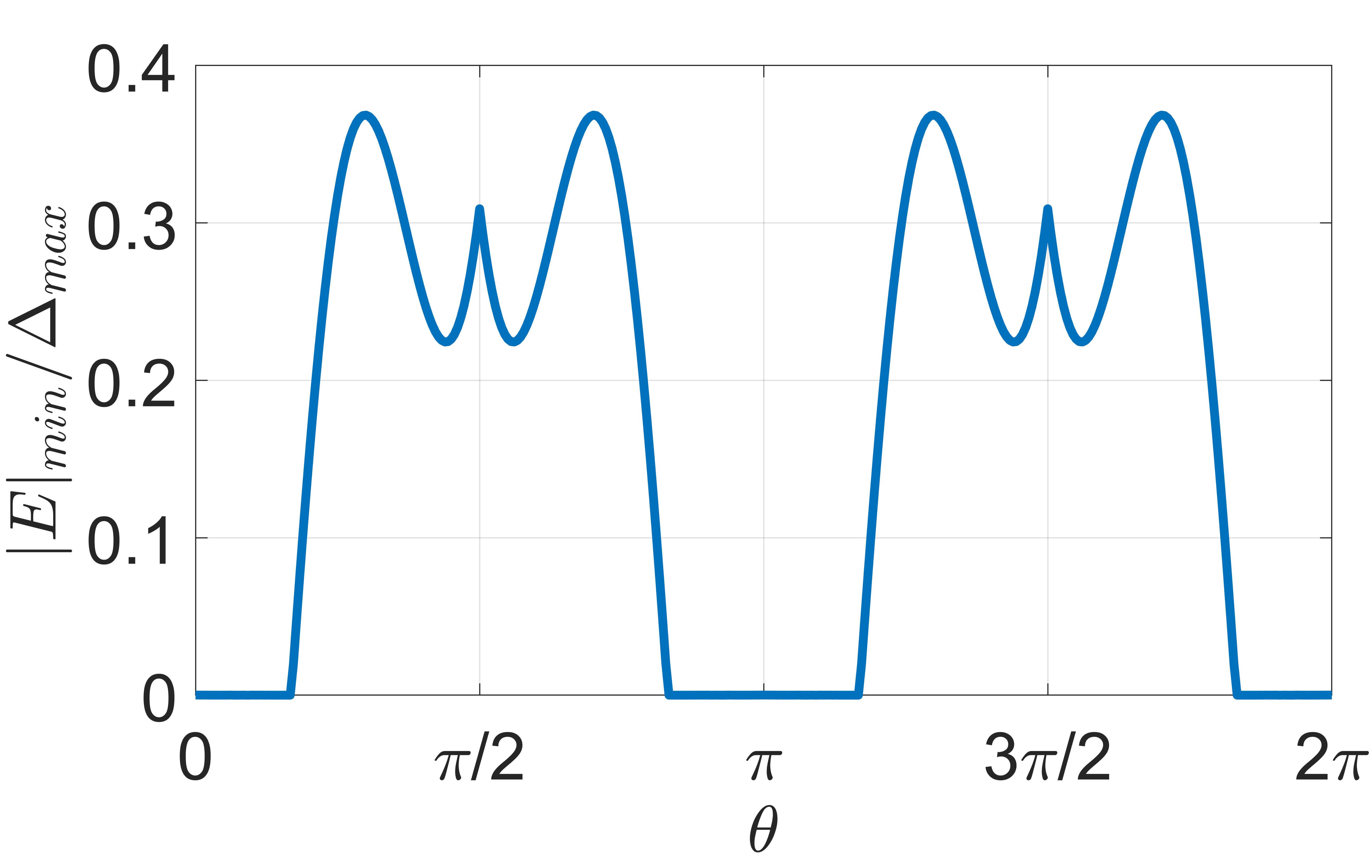}
\caption{Minimum single-particle energy gap as a function of polar angle predicted for FeSe$_{1-x}$S$_x$ near the $\Gamma$ point. For each angle $\theta$, the minimum energy gap occurs for a momentum approximately equal to $k_F$, the Fermi momentum.  The extended region of a zero energy gap is the predicted Bogoliubov Fermi surface.}
\label{MinE_PolarAngle} 
\end{figure}

The Bogoliubov quasiparticle spectrum depends upon the parity of the TRSB \cite{agterberg2017bogoliubov,brydon2018bogoliubov}. For even $\mathcal{P}$, it takes the form \cite{agterberg2017bogoliubov,brydon2018bogoliubov}
\begin{align}
E_{{\bf k},\pm,\nu}= \nu|{\bf h}({\bf k})|\pm\sqrt{[\xi_+({\bf k})
  -\mu]^2+|\psi({\bf k})|^2} ,
\label{Eeven}
\end{align}
where $\nu=\pm1$ and $\mu$ is the chemical potential.  For odd $\mathcal{P}$, this dispersion has been discussed  in Ref.~\cite{amin2024kramers} and takes the form
\begin{align}
E_{{\bf k},\pm}=\xi_-({\bf k}) \pm\sqrt{[\xi_+({\bf k})
  -\mu]^2+|\psi({\bf k})|^2}.
\label{Eodd}
\end{align}
For Eq.~(\ref{Eeven}), BFSs occur for ${\bf k}$ when $|{\bf h}({\bf k})|>|\psi({\bf k})|$, while for Eq.~(\ref{Eodd}) they occur for $|\xi_-({\bf k})|>|\psi({\bf k})|$. We note that for even $\mathcal{P}$, BFSs are topologically protected~\cite{agterberg2017bogoliubov, brydon2018bogoliubov}, while for odd $\mathcal{P}$, BFSs are not topologically protected, but they are still robust~\cite{tamm1932possible, timm2017inflated}. In Table~\ref{structures}, we show the BFSs that arise from Eqs.~(\ref{Eeven}) and~(\ref{Eodd}) for all irreps of interactions that break time-reversal symmetry. Note that we have chosen values of the parameters such that the spectrum contains BFSs. We have further chosen a finite value of $k_z$ when terms in either ${\bf h}({\bf k})$ or $\xi_-({\bf k})$ vanish by symmetry when $k_z=0$. For most of the irreps, the tetragonal symmetry is not broken. Importantly, the only irreps that allow the appearance of nematic BFSs are the $E_u$ and $E_g$ irreps. For these two representations, the structures of the BFSs are different. For the $E_u$ irrep, the nematic BFS with only two pockets along either the $k_x$ or $k_y$ direction is generic which agrees with the experiment~\cite{nagashima2022discovery}. Furthermore, the predicted minimum of the quasiparticle excitation energy as a function of polar angle, shown in Fig.~\ref{MinE_PolarAngle}, is in agreement with the laser ARPES measurements~\cite{nagashima2022discovery}. However, for the $E_g$ irrep, we generically find weakly nematic BFSs. In this case, when BFSs appear, there are typically four pockets with anisotropy in the size of these pockets along the $k_x$ and $k_y$ directions. This disagrees with the experimental observation for which only two pockets are seen. While it is possible that only two Fermi surfaces appear for the $E_g$ irrep, as we discuss next, this requires extreme fine-tuning and is highly unlikely. 

As Table~\ref{structures} shows, TRSB $E_g$ irrep interactions have two contributions. The first is the usual in-plane ferromagnetism described by the term $h_{1x}\hat{x}+h_{1y}\hat{y}$. The second contribution, $(h_{2x} k_x k_z+ h_{2y} k_y k_z)\hat{z}$, strongly varies with $k_z$ and is unlikely to be substantial in quasi-two-dimensional materials such as FeSe$_{1-x}$S$_x$. Without this second contribution, the usual ferromagnetic component yields Bogoliubov Fermi surfaces with tetragonal symmetry. The $k_z$-varying contribution is the origin of any nematic distortion of the BFSs in this case.  When this term is added and is small, there remain four BFSs with a weak nematicity as shown in Table~\ref{structures}. To reproduce the experimental observation of two nematic BFSs with the $E_g$ irrep requires the unphysical condition that the $k_z$-varying term dominates over the usual ferromagnetic component. We further note that ferromagnetism has not been observed in Fe-based superconductors and that density-functional theory (DFT) calculations show that ferromagnetism has an energy that is 68 meV/Fe higher than the DFT ground state, implying that ferromagnetism is completely unstable  ~\cite{mazin2023induced}. Based on the different predictions of the shape of BFSs for the $E_u$ and $E_g$ symmetries and the lack of experimental evidence for ferromagnetism, we conclude that the $E_g$ symmetry is vanishingly unlikely and the $E_u$ symmetry is the origin of the observed nematic BFS in the laser ARPES measurements.

\begin{table} 
\centering
\renewcommand{\arraystretch}{2.0}
\begin{tabular}{ | c | c | c | c| c| c |} 
\hline
Irrep & $\bf{h}({\bf k})$ & BFS   & Irrep & $\xi_-({\bf k})$ & BFS
\\ \hline
$A_{1g}$ & 
\makecell[c]{$h_1\left(k_y k_z \hat{x}-k_x k_z \hat{y}\right)$\\$+$$h_2 k_x k_y\left(k_x^2-k_y^2\right) \hat{z}$}  
     & \begin{minipage}[c]{0.08\columnwidth}
		\centering
		\raisebox{-.5\height}{\includegraphics[width=\linewidth]{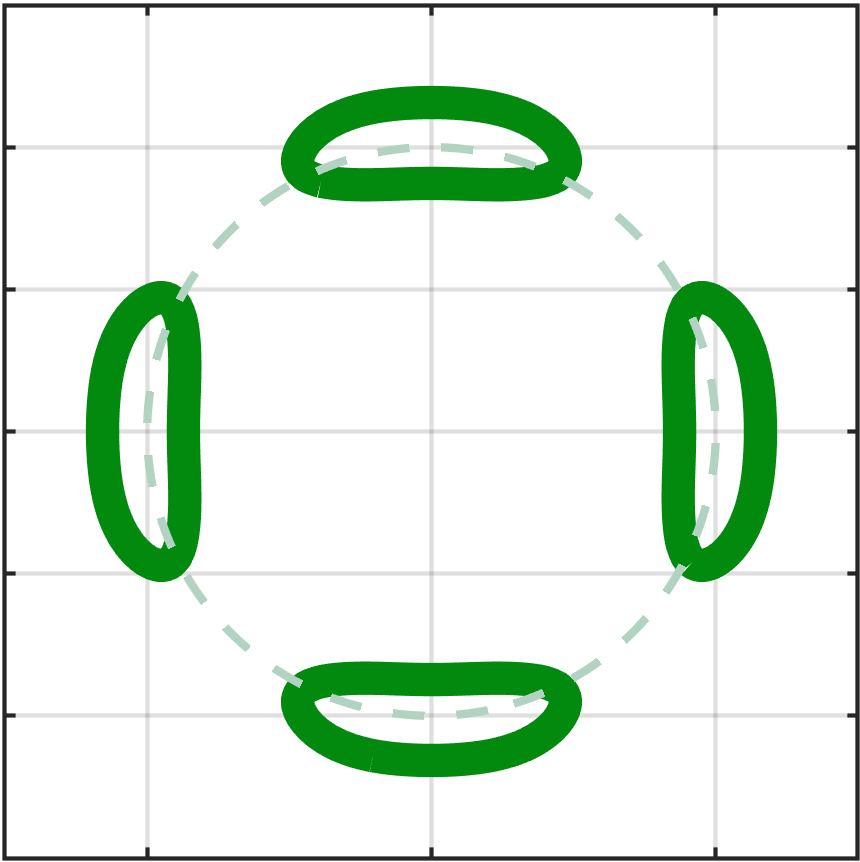}}
	\end{minipage}   
     & $A_{1u}$   
     & \makecell[c]{$\alpha k_x k_y k_z$ \\ $\times\left(k_x^2-k_y^2\right)$}
 &  \begin{minipage}[c]{0.08\columnwidth}
		\centering
		\raisebox{-.5\height}{\includegraphics[width=\linewidth]{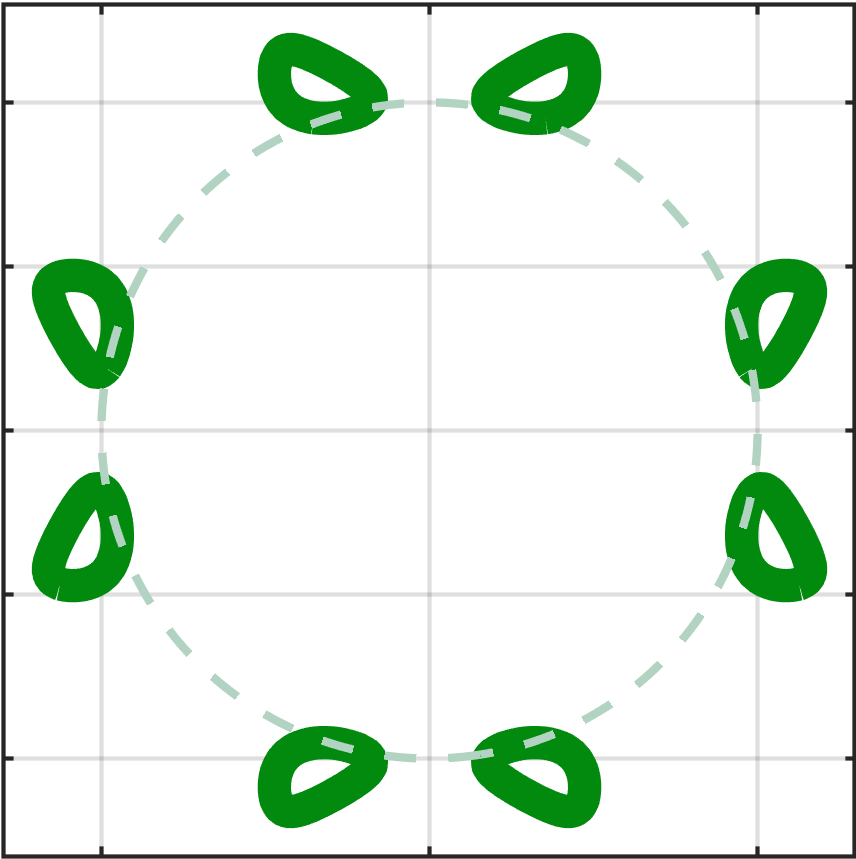}}
	\end{minipage}
\\ \hline
$A_{2g}$ & \makecell[c]{$h_1\left(k_x k_z \hat{x}+k_y k_z \hat{y}\right)$\\$ +$ $ h_2 \hat{z}$}   
    & \begin{minipage}[c]{0.08\columnwidth}
		\centering
		\raisebox{-.5\height}{\includegraphics[width=\linewidth]{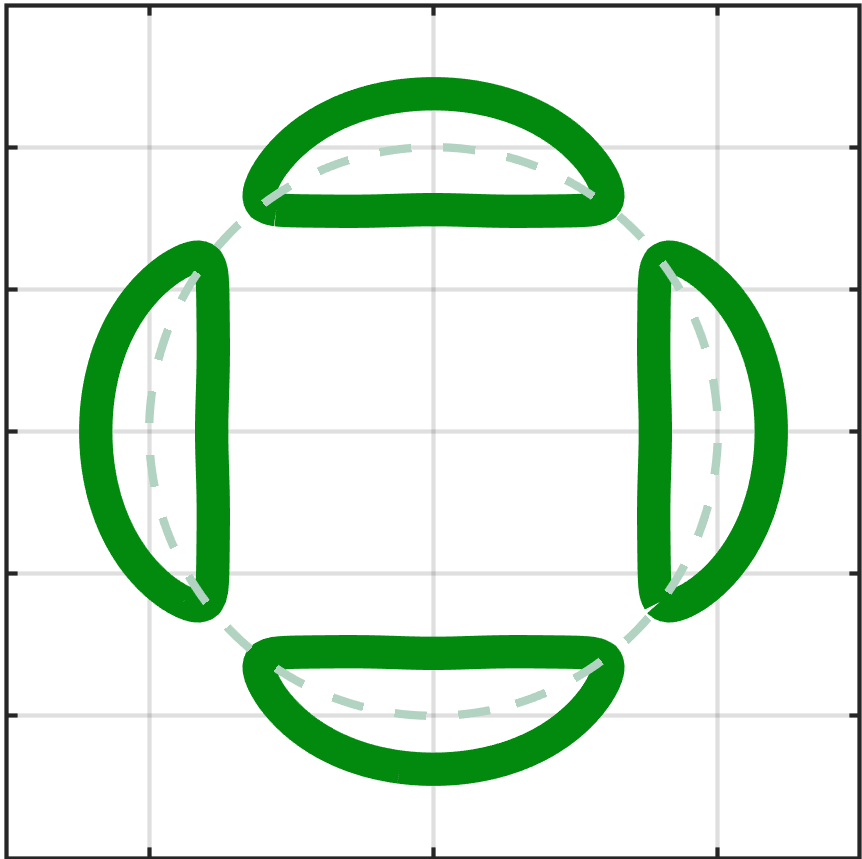}}
	\end{minipage}   
     & $A_{2u}$     
 & $\alpha k_z$  
 & \begin{minipage}[c]{0.08\columnwidth}
		\centering
		\raisebox{-.5\height}{\includegraphics[width=\linewidth]{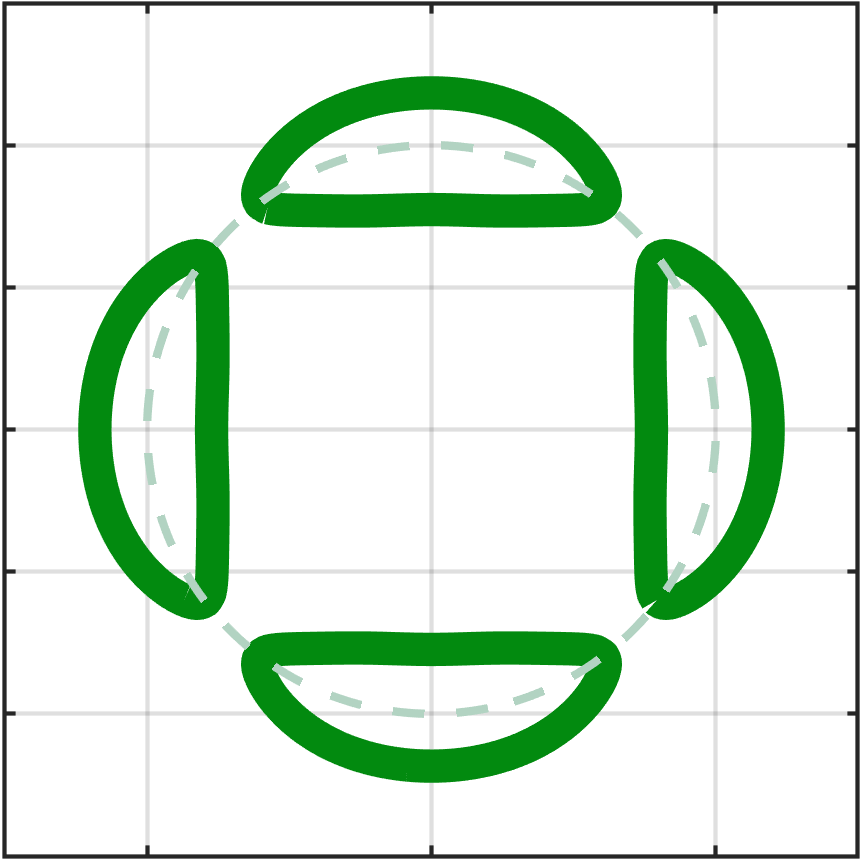}}
	\end{minipage}   
\\ \hline
$B_{1g}$ & \makecell[c]{$h_1\left(k_y k_z \hat{x}+k_x k_z \hat{y}\right)$\\$+$ $h_2 k_x k_y \hat{z}$} 
     & \begin{minipage}[c]{0.08\columnwidth}
		\centering
		\raisebox{-.5\height}{\includegraphics[width=\linewidth]{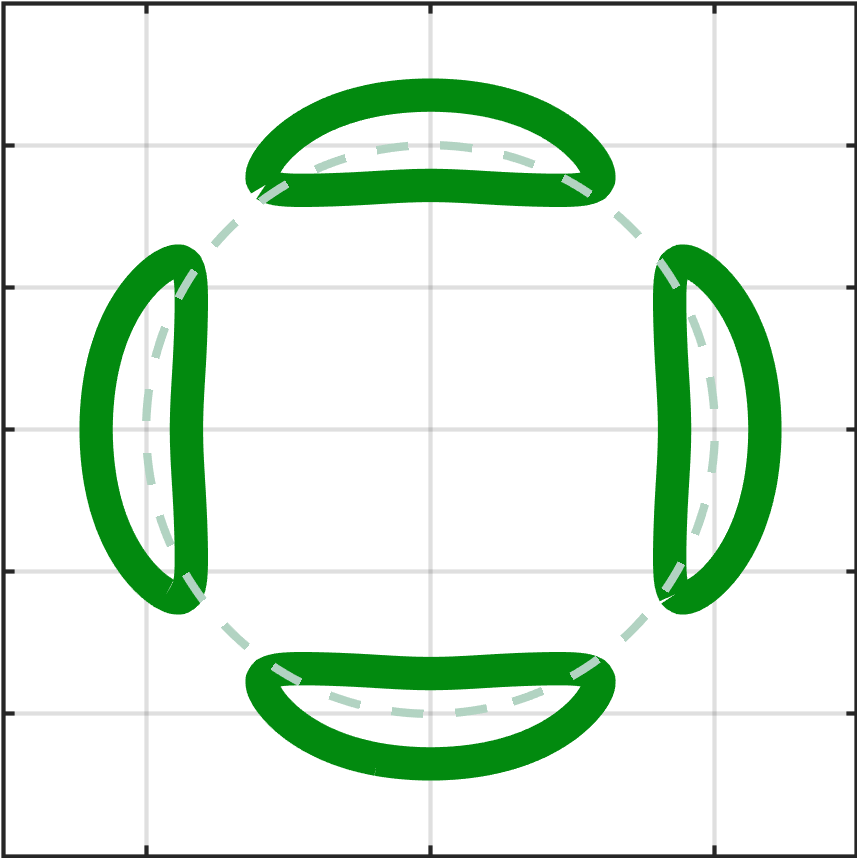}}
	\end{minipage}   
     & $B_{1u}$     
 & $\alpha k_x k_y k_z$  
 & \begin{minipage}[c]{0.08\columnwidth}
		\centering
		\raisebox{-.5\height}{\includegraphics[width=\linewidth]{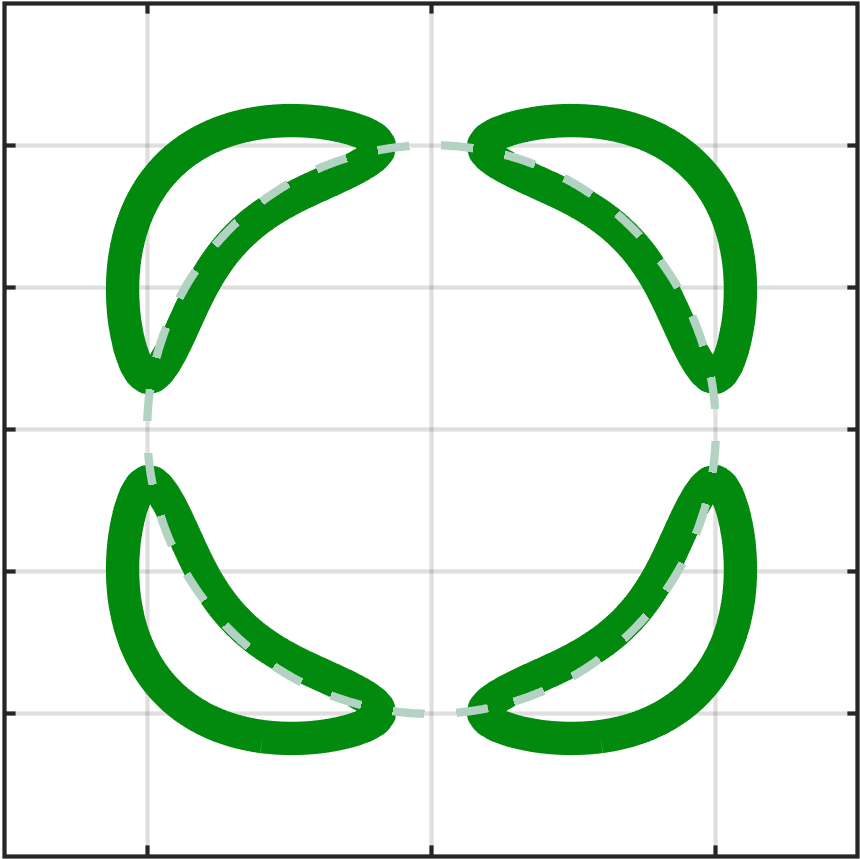}}
	\end{minipage}   
\\ \hline
$B_{2g}$ & \makecell[c]{$h_1\left(k_x k_z \hat{x}-k_y k_z \hat{y}\right)$\\$+$ $h_2\left(k_x^2-k_y^2\right) \hat{z}$} 
     & \begin{minipage}[c]{0.08\columnwidth}
		\centering
		\raisebox{-.5\height}{\includegraphics[width=\linewidth]{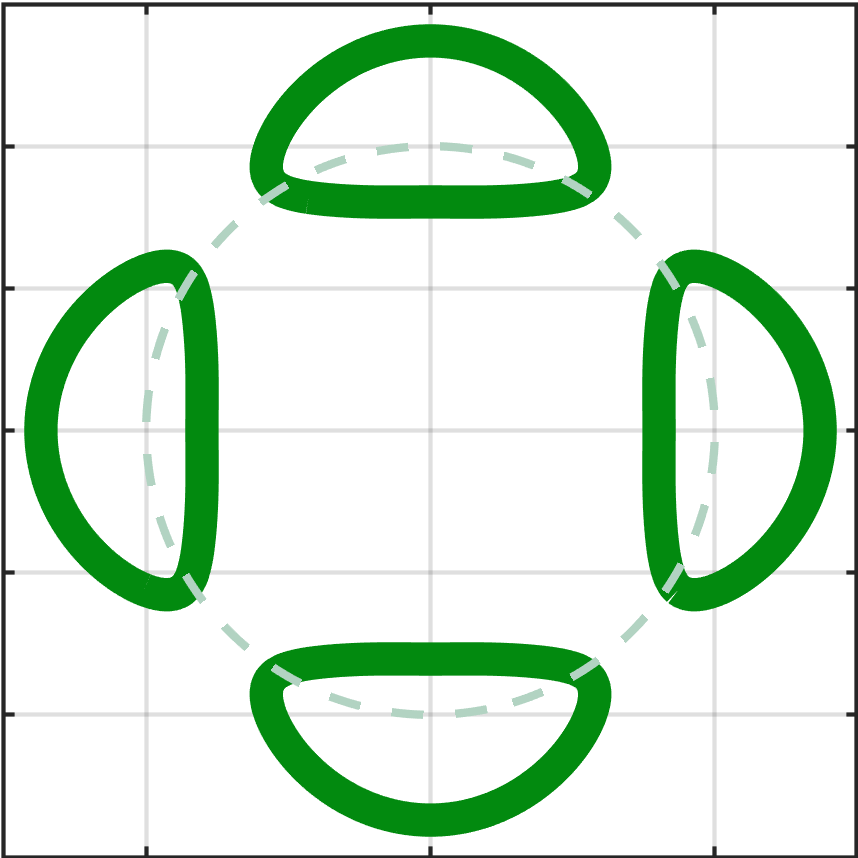}}
	\end{minipage}   
     & $B_{2u}$      
 & $\alpha k_z\left(k_x^2-k_y^2\right)$ 
 & \begin{minipage}[c]{0.08\columnwidth}
		\centering
		\raisebox{-.5\height}{\includegraphics[width=\linewidth]{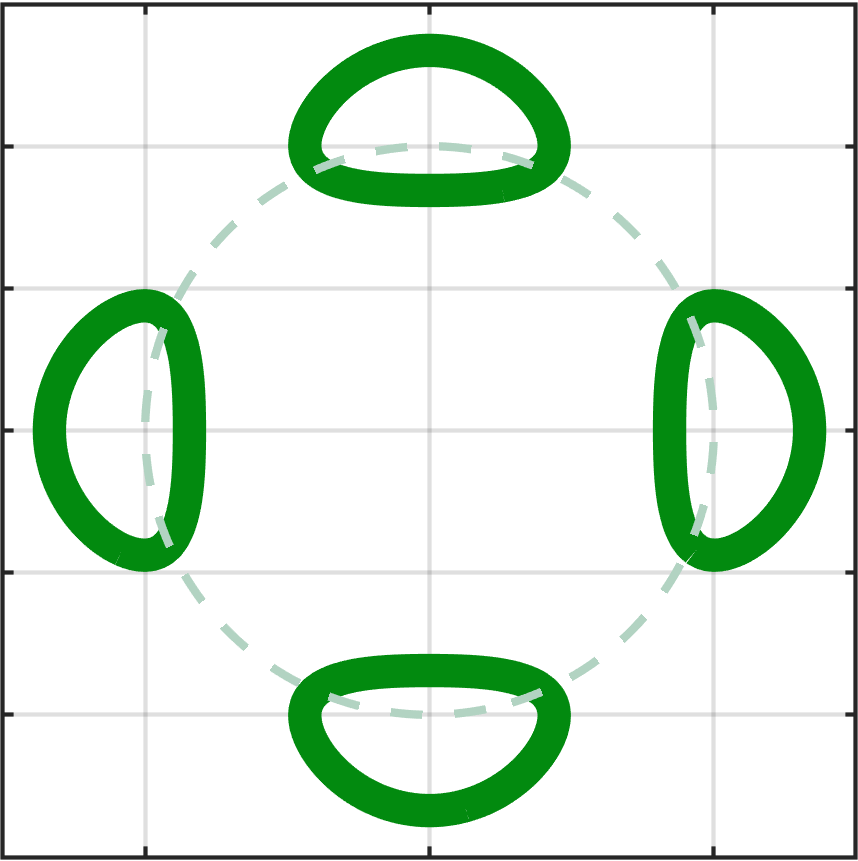}}
	\end{minipage} 
\\ \hline 
$E_{g}$ & \makecell[c]{$h_{1x}\hat{x}+h_{1y}\hat{y}$\\
$+$ $(h_{2x} k_x k_z$ \\
$+$ $h_{2y} k_y k_z)\hat{z}$}
     & \begin{minipage}[c]{0.08\columnwidth}
		\centering
		\raisebox{-.5\height}{\includegraphics[width=\linewidth]{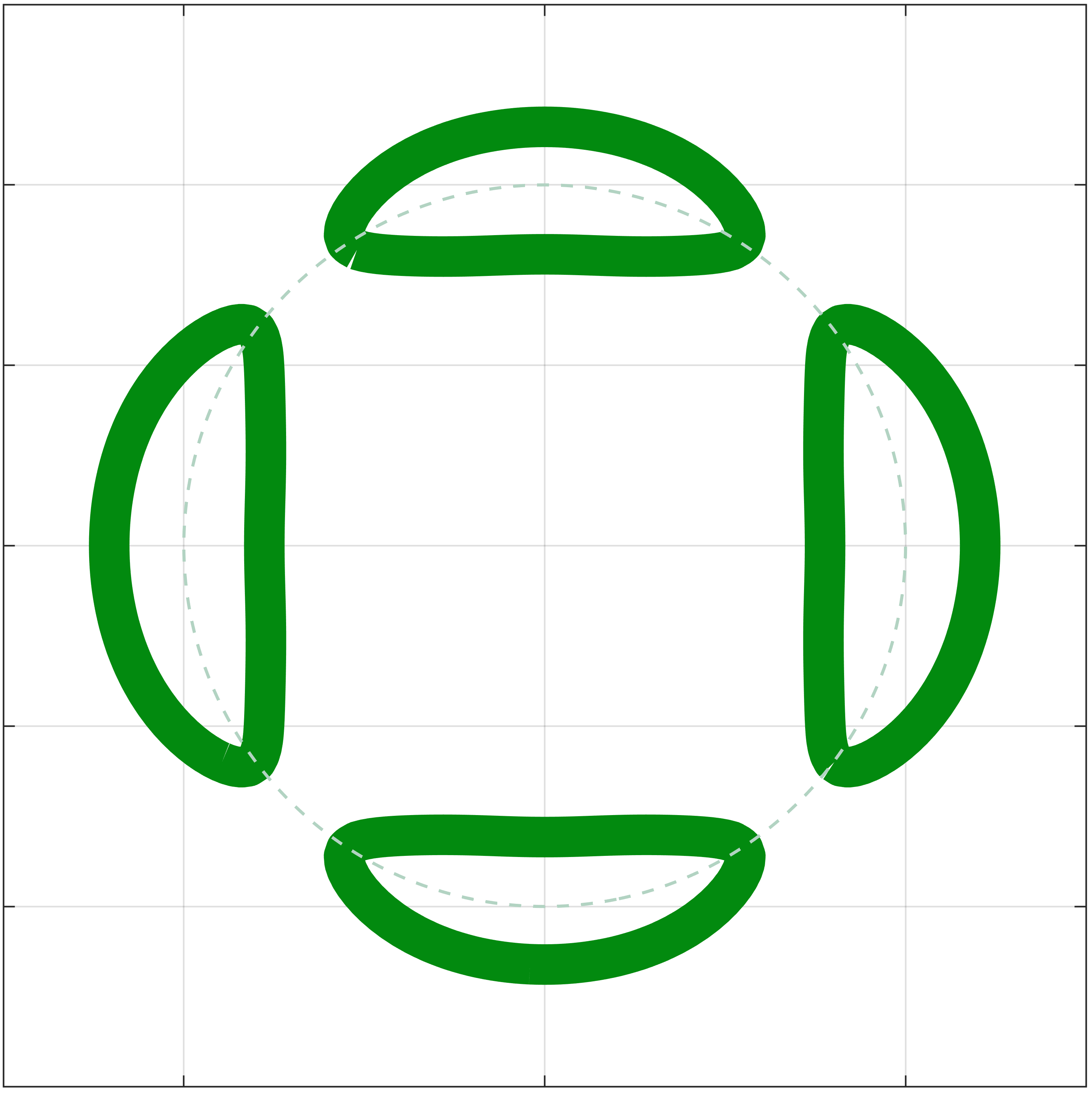}}
	\end{minipage}   
     & $E_{u}$   
 & $T_x k_x$ + $T_y k_y$  
 & \begin{minipage}[c]{0.08\columnwidth}
		\centering
		\raisebox{-.5\height}{\includegraphics[width=\linewidth]{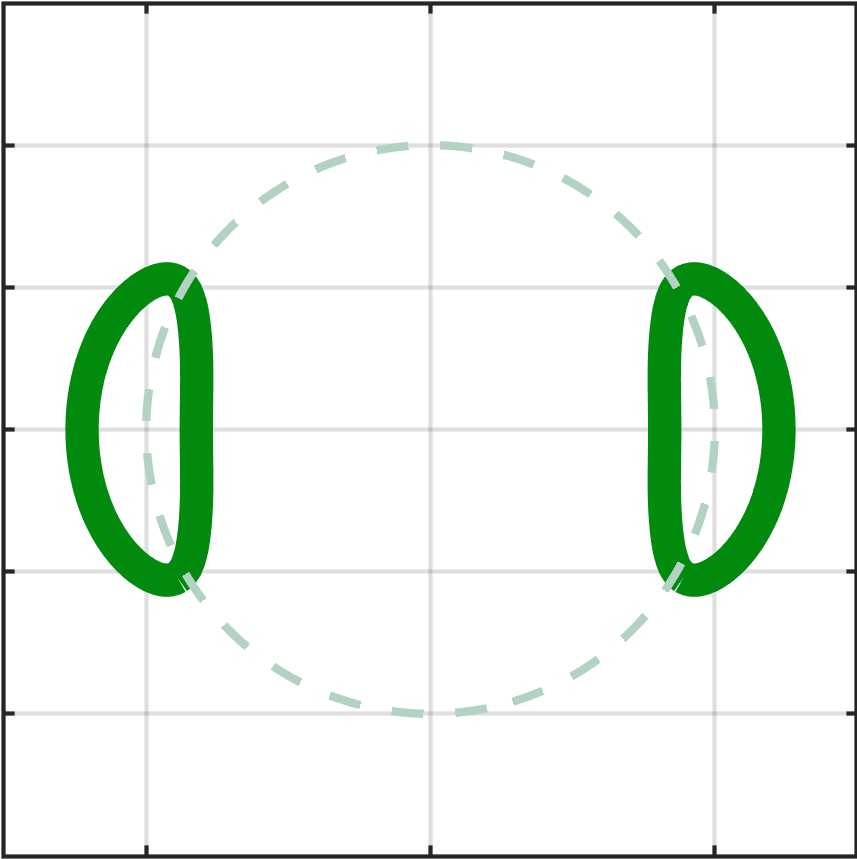}}
	\end{minipage} 
\\ \hline 
\end{tabular}
\caption{Structures of $\bf{h}({\bf k})$ and $\xi_-({\bf k})$ for $\mathcal{T}$-breaking order belonging to $D_{4h}$ point group irreps and the corresponding BFSs.  We express momenta $k$ in units of the Fermi momentum $k_F=\frac{\sqrt{2 m \mu}}{\hbar}$ and energies in units of the  maximum gap $\Delta_{max}=\Delta_0+\left|\Delta_4\right|$.  In units $\Delta_{max}$, we take $\mu=2.2361$, $\Delta_0=0.6545$, and $\Delta_4=-0.3455$. The other parameters in units $\Delta_{max}$ are chosen as: for $A_{1g}$, $h_1 k_z k_F=0.6325$, $h_2 k_F^4=0.7071$; for $A_{2g}$, $h_1 k_z k_F=0.3162$, $h_2=0.8944$; for $B_{1g}$, $h_1 k_z k_F=0.7746$, $h_2k_F^2=0.8944$; for $B_{2g}$, $h_1 k_z k_F=0.7746$, $h_2k_F^2=0.8944$; for $E_{g}$, $h_{1x}=0.8367$, $h_{1y}=0$, $h_{2x}k_zk_F=0.5477$, $h_{2y}k_zk_F=0$; for $A_{1u}$, $\alpha k_z k_F^4=2.3$; for $A_{2u}$, $\alpha k_z =0.95$; for $B_{1u}$, $\alpha k_z k_F^2=2.2$; for $B_{2u}$, $\alpha k_z k_F^2=1$; for $E_{u}$, $T_x k_F=0.95$ and $T_y k_F=0 $.}
\label{structures}
\end{table}

{\it Origin of magnetic toroidal order.}---The above symmetry-based analysis suggests that the TRSB that gives rise to the observed nematic BFSs is a magnetic toroidal order belonging to an $E_u$ irrep. Here we show two possible microscopic origins for this order realizing the $E_u$ symmetry: N\'{e}el AFM order with moments oriented in plane or the spontaneous formation of a PDW phase. 

There is plenty of experimental evidence suggesting that the AFM order is in the family of iron-based superconductors including FeSe ~\cite{dai2012magnetism, ding2014antiferromagnetism, dai2015antiferromagnetic, biswal2021recent}. In inelastic neutron-scattering experiments, it was reported that there was the coexistence of N\'{e}el AFM fluctuations and stripe AFM fluctuations in FeSe~\cite{wang2016magnetic}. Once FeSe enters the tetragonal phase, N\'{e}el AFM fluctuations are more important and are clearly observed. These neutron results are consistent with the explanation of the Raman spectra of FeSe compared to simulations of a frustrated spin-1 system~\cite{baum2019frustrated}. There was also experimental evidence showing the in-plane checkerboard AFM order in monolayer FeSe~\cite{qiao2020fingerprint}. 

In addition, it has been shown by DFT calculations that the checkerboard AFM order is a realistic ground state in monolayer FeSe~\cite{shishidou2018magnetic}. Furthermore, in a recent experiment-driven theory~\cite{mazin2023induced}, the authors propose that the checkerboard AFM order is a natural explanation for TRSB observed in FeSe/SrTiO$_3$ heterostructures~\cite{zakeri2023direct} and in the surface layer of superconducting Fe(Se, Te) compositions~\cite{farhang2023revealing}. 

While evidence for PDW order in Fe-based superconductors is not as strong as that for checkerboard magnetic order, there has been an experimental observation of the PDW state in monolayer Fe(Te, Se) films grown on SrTiO$_3$(001) substrates~\cite{liu2023pair}.

\begin{figure} 
\centering
\includegraphics[height=7cm]{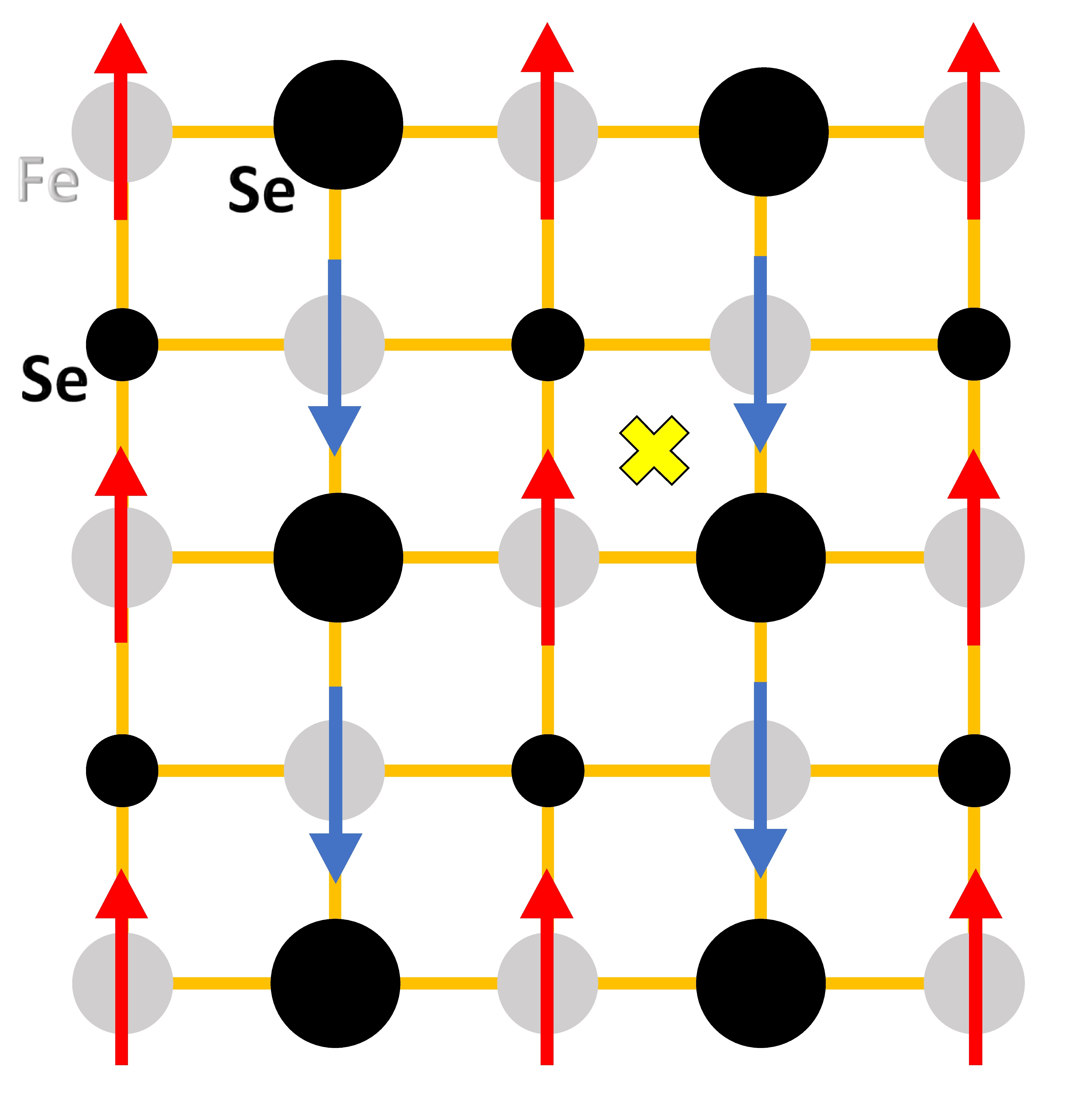}
\caption{N\'{e}el antiferromagnetic order that serves as a magnetic toroidal order. The selenium atoms with bigger (smaller) sizes reside above (below) the iron layer. The yellow cross is the lattice inversion center. The red and blue arrows indicate the direction of the magnetic moments at the Fe sites.}
\label{AFM}
\end{figure}

We first discuss the N\'{e}el AFM order. In Fig.~\ref{AFM} we show the N\'{e}el AFM order that belongs to the $E_u$ irrep. Key to understanding how this N\'{e}el ordered state is odd under parity symmetry is the nonsymmorphic P4/nmm space group of tetragonal FeSe$_{1-x}$S$_x$. This space group requires that there are two Fe atoms per unit cell related by an inversion center. Consequently, when the moments on these two Fe sites are oriented in opposite directions, the magnetic state is odd under $\mathcal{P}$. This, together with being odd under $\mathcal{T}$, illustrates that this is a magnetic toroidal order parameter.

To understand how this AFM order gives rise to the term  $\xi_{-}({\bf k})$ in Eq.~(\ref{singleband}), it is useful to consider a simple tight-binding model for FeSe$_{1-x}$S$_x$. In particular, consider a 2D model that only includes $xy$ orbitals on the iron (Fe) sites.  The corresponding tight-binding Hamiltonian is
\begin{align}
H_0 = 
  \ &[t_1(\cos k_x+\cos k_y)-\mu]\tau_0\sigma_0+t_{2}\cos \frac{k_x}{2}\cos\frac{k_y}{2}\tau_x\sigma_0 
\notag\\
&+\alpha_R(\sin k_x \tau_z\sigma_y+\sin k_y \tau_z\sigma_x),
 &\label{eq:H0}
\end{align}
where the $\alpha_R$ term is a Rashba-like spin-orbit coupling \cite{agterberg2017resilient}. We add to $H_0$ the N\'{e}el AFM order with moments oriented along the ${\hat y}$ direction. This is given by $T_x \tau_z\sigma_y$. Treating this as a perturbation to the Hamiltonian $H_0$ yields 
\begin{align}
\label{epsilon_odd}
\xi_{-}({\bf k})&=\frac{T_x\alpha_R\sin k_x}{\sqrt{ t_2^2(\cos \frac{k_x}{2}\cos\frac{k_y}{2})^2+ \alpha_R^2(\sin^2 k_x+\sin^2 k_y)}}.
\end{align}

{\it Lifshitz invariant and PDW state.}---The simultaneous breaking of time-reversal symmetry and parity symmetry by the magnetic toroidal order allows the existence of a Lifshitz invariant in the Ginzburg-Landau free energy. Both the magnetic toroidal order $\boldsymbol{T}=\left(T_x, T_y\right)$ and $\boldsymbol{D}=\left(D_x, D_y\right)$ (where $\boldsymbol{D}=-i{\nabla}-2 e \boldsymbol{A}$, the charge of the electron $e<0$, $\boldsymbol{A}$ is the vector potential, and we work in units such that $\hbar=c=1$) transform as an $E_u$ irrep of $D_{4 h}$. Therefore, these can be coupled leading to the Lifshitz invariant
\begin{equation}
\epsilon \boldsymbol{T}\cdot\left[\psi\left( \boldsymbol{D} \psi\right)^*+\psi^*\left( \boldsymbol{D} \psi\right)\right].
\label{free_energy}
\end{equation}
%
This Lifshitz invariant guarantees 
a PDW state with finite momentum 
pairing~\cite{mineev1994helical,Smidman:2017}, in which the superconducting 
order parameter develops a spatial variation, $\psi=\psi_0 e^{i 
\boldsymbol{q} \cdot \boldsymbol{r}}$. As discussed in Ref.~\cite{amin2024kramers}, the PDW state with 
finite $\boldsymbol{q}$ alters the Bogoliubov quasiparticle spectrum:
\begin{equation}
E_{{\bf{k}}, \boldsymbol{q}, \pm}=\xi_{-}({\bf{k})}+\frac{\hbar}{2} \boldsymbol{q} \cdot \boldsymbol{v}_F \pm \sqrt{\left [\xi_+(\bf{k})-\mu \right]^2+|\psi(\bf{k})|^2},
\label{EqBogo}
\end{equation}
where $\boldsymbol{v}_F$ is the Fermi velocity. As shown in Ref.~\cite{amin2024kramers}, the main consequence of $\boldsymbol{q}$ is to reduce the size of the BFSs. However, the BFSs are generically not fully removed by a nonzero $\boldsymbol{q}$.

{\it Superconducting fluctuations and AFM $T_c$.}--- 
For the N\'{e}el AFM order to be the origin of the BFSs, it needs to order above the superconducting $T_c$.  We suggest that the observed $\mu$SR signal \cite{ishida2022pure} onsets due to the formation of AFM order that occurs just above the superconducting $T_c$. This may occur if there is an enhancement of the $T_c$ of the AFM order by superconducting fluctuations.
To illustrate this effect, we consider the following minimal free energy density:
	\begin{equation}
		f=\frac{\kappa}{2}|\nabla\psi|^2+\frac{\mu}{2}|\psi|^2+\frac{\alpha_T}{2}|{\bf T}|^2 
        +\epsilon{\bf T}\cdot\left[ \psi (-i{\bf \nabla} \psi)^*+ \mathrm{c.c.}\right].
        \label{fminimal}
	\end{equation}
	By integrating out the superconducting fluctuations, we obtain the correction to $\alpha_T$:
	\begin{align}
		&\alpha_T\rightarrow \mu_T= \alpha_T-\frac{1}{2T}\epsilon^2\langle |\psi|^2 |\nabla\psi|^2 \rangle \notag \\
         &=\alpha_T-2T\epsilon^2\left(\frac{1}{\Lambda^3}\int^{\Lambda^3} \frac{d^3k}{(2\pi)^3}\frac{1}{\mu+\kappa {\bf k}^2}\right)\notag\\
         &\times \left(\frac{1}{\Lambda^3}\int^{\Lambda^3} \frac{d^3k}{(2\pi)^3}\frac{k^2}{\mu+\kappa {\bf k}^2}\right),
	\end{align}
	where $\Lambda$ is the ultraviolet wavelength cutoff. Near the superconducting transition ($\mu \rightarrow 0$), the fluctuations of the superconducting order parameter, $\langle |\psi|^2\rangle$ and $\langle |\nabla\psi|^2 \rangle$, reach their maximum values allowing the magnetic toroidal order to appear at $\mu_T=0$. We note that these materials host strong superconducting fluctuations \cite{kasahara2016giant} consistent with such a possibility of a superconductivity-driven AFM order.

{\it Spontaneous PDW order.}---While the above explanation offers a mechanism for the proximity of the superconducting and the AFM $T_c$, Eq.~(\ref{EqBogo}) suggests a mechanism for which there is only a single $T_c$ at which superconductivity and AFM appear together with BFSs. In particular,  this will occur if a spontaneous PDW order appears at $T_c$ (for which $\psi=\psi_0 e^{i 
\boldsymbol{q} \cdot \boldsymbol{r}}$). For this to occur, the stiffness $\kappa$ in Eq.~(\ref{fminimal}) needs to be less than zero, $\kappa<0$. While uncommon, there are two mechanisms that could allow this to occur. The first is in a multiband system, where in addition to the usual positive single-band stiffness, there can be negative contributions to the stiffness that arise from quantum geometry~\cite{kitamura2022quantum, jiang2023pair,chen2022pair}. The second is that AFM fluctuations can reduce $\kappa$, much like superconducting fluctuations reduced $\alpha_T$~\cite{setty2023mechanism}.

{\it Discussion.}---There is one additional consequence of the existence of the Lifshitz invariant in Eq.~(\ref{free_energy}). If we have a current-carrying state in the superconductor, $\boldsymbol{q}$ becomes nonzero. Since $\boldsymbol{q}$ and $\boldsymbol{T}$ are bilinearly coupled, $\boldsymbol{T}$ will be nonzero. Therefore, we conclude that supercurrents will induce N\'{e}el magnetic order in many Fe-based superconductors that have P4/nmm space group symmetry and two Fe ions per unit cell. We can estimate the order of magnitude of the magnetization $M \sim \frac{\chi}{\chi_0} \frac{\alpha_R}{t_2} N(0) \Delta \mu_B$, where $\chi$ is the full susceptibility, $\chi_0$ is approximately $N(0) \mu_B^2$, $\alpha_R$ and $t_2$ are parameters we introduced in Eq.~(\ref{eq:H0}). We also point out that with the checkerboard antiferromagnetic order, we expect the superconducting critical temperature $T_c$ to be lower than without this magnetic order due to the existence of BFSs. This may account for the observed decrease in $T_c$~\cite{mizukami2023unusual} We further note that the existence of the finite momentum pairing is crucial for the superconducting diode effect \cite{yuan2022supercurrent, he2022phenomenological, daido2022intrinsic, pal2022josephson}. This suggests that superconductivity coexisting with the magnetic toroidal order provides a route toward creating the superconducting diode effect.

Here we have assumed an $s$-wave gap function. Generically, for any $s$-wave and $p$-wave gap functions, the same results for the BFSs hold. However, for $d$-wave gap functions, the predicted anisotropy of the BFSs will be different, and this is not in agreement with the laser ARPES measurements.

{\it Conclusions.}---We performed a symmetry-based analysis of BFSs that can arise from TRSB for FeSe$_{1-x}$S$_x$. We have shown that the origin of TRSB and the nematic BFS in the tetragonal phase is a magnetic toroidal order belonging to the E$_u$ representation.  We are able to replicate the BFS shape and the minimum quasiparticle excitation energy observed by the laser ARPES measurements. We point to two possible origins of the MT order, either through static N\'{e}el  AFM order or due to the spontaneous formation of PDW superconductivity. We argue that supercurrents will induce N\'{e}el magnetic order in many Fe-based superconductors.

Recently, we learned that an alternate scenario for the phenomenology of
tetragonal FeSe$_{1-x}$S$_x$ has been investigated within a microscopic model with magnetic interactions by Yifu Cao, Chandan Setty, Laura Fanfarillo, Andreas Kreisel, and P. J. Hirschfeld~\cite{cao2023microscopic}.

\vglue 0.5 cm
\begin{acknowledgments} 
D.F.A., H.W., A.A., and Y.Y. were supported by the U.S. Department of Energy, Office of Basic Energy Sciences, Division of Materials Sciences and Engineering under Award No. DE-SC0021971. We acknowledge useful discussions with Rafael Fernandes, Peter Hirschfeld, Takasada Shibauchi, Christian Parsons, and Amalia Coldea.

\end{acknowledgments}

\bibliography{biblio}

\end{document}